\hsize=16.6truecm        \hoffset=-0.3truecm      \vsize=25truecm
\overfullrule=0pt        \parindent0.7true cm     \voffset0.0truecm
\tolerance=250           \raggedbottom
\font\rm=cmr10           	  \font\math=cmsy10
\font\it=cmti10                    
\font\gk=cmmi10          \font\smtt=cmtt10 at 5pt \font\sssrm=cmr10 at 5pt
\font\brm=cmr10 scaled\magstep1          
\font\bit=cmti10 scaled\magstep1          
         \font\ssmmath=cmsy10 at 6pt
         \font\ssrm=cmr10 at 6pt
         
\font\biggk=cmmi10 at 12pt               \font\smgk=cmmi10 at 8pt
\font\smmath=cmsy10 at 8pt               \font\sssmmath=cmsy10 at 4pt
\font\bigmath=cmsy10 at 12pt             \font\ssit=cmti10 at 7pt
\font\srm=cmr10 at 8pt   \font\sit=cmti10 at 8pt  \font\sbf=cmbx10 at 8pt
\font\stmath=cmsy10 at 7pt               \font\stgk=cmmi10 at 7pt
\def\cl{\centerline}     \def\ni{\noindent}       
\def\bs{\bigskip}        \def\ms{\medskip}	  
\def\vs#1{\vskip#1pt}    \def\hs#1{\hskip#1pt}    \def\hm#1{\hskip#1em}
\def\vfe{\vfill\eject}   
\def\hf{\hfill}          \def\hah{\hf&\hf}
\def\hcr{\hf\cr}            
  \def\PM{\hs1 $\pm$ \hs1} \def\PN{$\pm$ }

\def\l1{\looseness-1}    
 \def\viz{{\it viz.,}\hs3}   \def\ie{{\it i.e.}\hs2}
\def\msp{\kern.4em}        \def\={= \sk}
\def\kms{km~s$^{-1}$}    \def\kmss{\kms\hs2}         
\def\bigkms{\hbox{km s\raise.8ex\hbox{\rm\hs{0.8}--\hs{0.5}1}}}
\def\smkms{\hbox{km s\raise.6ex\hbox{\ssrm\hs{0.8}--\hs{0.5}1}}}
           
\def\itkms{\hbox{{\it km s}\raise.6ex\hbox{\sit\hs{0.8}--\hs{0.5}1}}\hs1}
\def\sitkms{\hbox{{\sit km s}\rlap{\raise.6ex\hbox{\ssit\hs{0.8}--\hs{0.5}1}}\hs4}}
\def\itamm{\hbox{{\it \AA~mm}\raise.7ex\hbox{\sit\hs{0.8}--\hs{0.5}1}}\hs1}
\def\up#1{\raise.9ex\hbox{\ssrm#1}}     \def\bsk#1{\baselineskip#1pt}
\def\upnormal#1{\hs{0.5}\raise.9ex\hbox{\srm#1}\hs3}
\def\down#1{\lower.5ex\hbox{\ssrm#1}\hs3}          \def\:{:\hs3}
\def\downormal#1{\lower.5ex\hbox{\srm#1}\hs3}
\def\downit#1{\lower.5ex\hbox{\ssit#1}\hs3}
                 \def\sec{\hbox{$''$\hs{-2}.}}  
\def\bigsec{{\raise.9ex\hbox{{\smmath\char'60\char'60}}}\hs{-1}.}
\def\smsec{\hs1\raise1.0ex\hbox{{\sssmmath\char'060\hs{0.5}\char'060}}.}
\def\smsecs{\hs1\raise1.0ex\hbox{{\sssmmath\char'060\hs{0.5}\char'060}}\hs2}
\def\bigsecs{{\raise.9ex\hbox{{\smmath\char'60\char'60}}}\hs2}
\def\ea{{\it \hbox{et al.}}}         \def\eas{\ea\hs3}
\def\smea{{\sit et al}.}            \def\smeas{\smea\hs3}
\def\etal#1{\hbox{{\it et al\/}.\hs{-0.8}$^{#1}$}}  
\def\setal{{\sit et al\/}.,\hs3}

\def\amm{\AA~mm$^{-1}$}             \def\amms{\AA~mm$^{-1}$\hs3}
    
\def\ahg{\amms at H$\gamma$}        
        
\def\msun{$M\hs{-1.5}_{\odot}$}     \def\msuns{\msun\hs3}
\def\bigmsun{{\bit M}\lower0.5ex\hbox{\math\char'14}} 
\def\smmsun{{\sit M}\lower0.5ex\hbox{\ssmmath\char'14}}

\def\dego{\raise.9ex\hbox{\smmath\char'16}}     
\def\smdego{\raise.9ex\hbox{\ssmmath\char'16}}  
\def\bigdego{\raise.9ex\hbox{\math\char'16}}    
\def\degs{\dego\hs3}                               
\def\b{\hm{0.5}}  \def\k{\hm{0.3}}   \def\sk{\hm{0.6}}
          
\def\M{\raise.9ex\hbox{\srm m}}      
\def\bigM{\raise.9ex\hbox{\rm m}}     
\def\mag{\hbox{\raise.9ex\hbox{\srm m}\hs{-1}.}}      
\def\bigmag{\raise.9ex\hbox{\rm m}\hs{-1}.}
\def\smmag{\raise.9ex\hbox{\ssrm m}\hs{-1}.}
\def\dmv{\hs{-1}{\gk\char'1}{\it m}\lower0.5ex\hbox{{\sit V}}}

\def\smdmv{\hs{-1}{\smgk\char'1}{\sit m}\lower0.5ex\hbox{{\ssit V}}}
\def\dMv{\hs{-1}{\gk\char'1}{\it M}\lower0.5ex\hbox{{\sit V}}}

\def\2smlambda{\hs{-1}{\smgk\char'25\char'25}}  \def\2smlambdas{\2smlambda\hs3}
  \def\K2{\hs{-1}{\it K}\hs{-1}$_2$}
\def\smK#1{{\sit K\lower0.5ex\hbox{\ssrm #1}}}
\def\smq{{\sit q}\hs9\srm (= {\sit m}\lower0.5ex\hbox{\ssrm 1}/{\sit m}\lower0.5ex\hbox{\ssrm 2})}

\def\asini{\hbox{$a_1$\hs1sin\hs2$i$}}          
          
\def\sma#1sini{{\sit a\/}\lower0.5ex\hbox{\ssrm #1}{\srm\hs2  sin}\hs1{\sit i}}
\def\smasini{{\sit a}\lower0.5ex\hbox{{\ssrm 1}}\hs2 sin\hs2{\sit i} \hs3 (Gm)}
\def\vsini{\hbox{$v$\hs{1.5}sin\hs{1.5}$i$}}   \def\vsinis{\vsini\hs4}

\def\smfm#1{{\sit f\/}({\sit m\/}\lower0.5ex\hbox{\ssrm #1}) \hs8 ({\sit M}\lower0.5ex\hbox{\ssmmath\char'14})}
\def\smm#1sin3i{{\sit m\/}\lower0.5ex\hbox{\ssrm #1}\srm\hs1sin\raise0.8ex\hbox{\ssrm 3}\hs1{\sit i} \hs{1.2}({\sit M}\lower0.5ex\hbox{\ssmmath\char'14})}

\def\velocity{\hbox{velocity}\hs3}      
\def\velocities{\hbox{velocities}\hs3}          \widowpenalty=1000
\def\zmsin3i{$m_1$\hs1 sin$^3$\hs1 $i$}         \hyphenpenalty=200
\def\m2sin3i{$m_2$\hs1 sin$^3$\hs1 $i$}         \clubpenalty=1000
                 
                                \brokenpenalty=1000
\def\gD{\hbox{$\gamma$ Dorad\^us}}              \def\gDs{\gD\hs3}

\def\cor{\hbox{{\it C\hs{-1}oravel\/}}}         \def\cors{\cor\hs4}
 
\def\smcor{\hbox{\sit C\hs{-0.5}oravel\/}}      \def\smcors{\smcor\hs4}
     
\def\b-v{\hbox{($B-V$)}\hs4}                    \def\u-b{\hbox{($U-B$)}\hs4}  
\def\smb-v{({\sit B {\smmath\char'0} V\/})}     
\def\bigb-v{(\hs{-1}{\bit B {\bigmath\char'0}\hs4V\/})\hs4}
\def\smu-b{({\sit U {\smmath\char'0} B\/})}
\def\dmc0{\hbox{{\gk\char'1}\hs{-0.5}$M_{c_0}$\hs0}}
\def\smdmc0{{\smgk\char'1}{\sit M\lower0.5ex\hbox{c\/\lower0.5ex\hbox{{\sssrm 0}}}}}
\def\frac#1/#2{\leavevmode\kern.05em\raise.6ex\hbox{\the\scriptfont0
    #1}\kern-.15em/\kern-.15em\lower.3ex\hbox{\the\scriptfont0 #2}\hs1}
\def\fracs#1/#2{\leavevmode\kern.05em\raise.6ex\hbox{\the\scriptfont0
    #1}\kern-.15em/\kern-.15em\lower.3ex\hbox{\the\scriptfont0 #2}\hs4}
\def\smfrac#1/#2{\leavevmode\kern.05em\raise.6ex\hbox{\the\scriptscriptfont0
    #1}\kern-.15em\raise.1ex\hbox{{\srm /}}\kern-.15em\lower.3ex
    \hbox{\the\scriptscriptfont0 #2}\hs1}
\def\smfracs#1/#2{\leavevmode\kern.05em\raise.6ex\hbox{\the\scriptscriptfont0
    #1}\kern-.15em\raise.1ex\hbox{{\srm /}}\kern-.15em\lower.3ex
    \hbox{\the\scriptscriptfont0 #2}\hs3}
\def\bigfrac#1/#2{\leavevmode\kern.05em\raise.6ex\hbox{\rm
    #1}\kern-.2em\raise.1ex\hbox{\brm /}\kern-.15em\lower.3ex\hbox{\rm #2}\hs1}
\def\bigfracs#1/#2{\leavevmode\kern.05em\raise.6ex\hbox{\rm
    #1}\kern-.2em\raise.1ex\hbox{\brm /}\kern-.15em\lower.3ex\hbox{\rm #2}\hs4}
\def\gsim{{\hbox{\raise.5ex\hbox{$>$}}\hskip-.8em\lower.6ex\hbox{$\sim$}}\hs3}
\def\lsim{{\hbox{\raise.4ex\hbox{$<$}}\hskip-.8em\lower.5ex\hbox{$\sim$}}\hs3}
\def\smgsim{{\hbox{\raise.5ex\hbox{{\stgk\char'76}}}\hskip-.7em\lower.5ex\hbox{{\stmath\char'30}}}\hs2}
\def\smlsim{{\hbox{\raise.5ex\hbox{{\stgk\char'74}}}\hskip-.7em\lower.5ex\hbox{{\stmath\char'30}}}\hs2}
\def\biglsim{{\hbox{\raise.5ex\hbox{{\biggk\char'74}}}\hskip-.7em\lower.5ex\hbox{{\bigmath\char'30}}}\hs2}
\def\g{\hbox{$\gamma$-velocity}}                  \def\gs{\g\hs3}

\def\pip{{\hs{-1}\ssmmath\hs2\raise.5ex\hbox{\char'17}}}  
  
\def\section#1{\bs\ni\it#1\ms\rm}
\def\*#1{{\bsk{12}{\parindent9pt\footnote*{\hs{-5}{\srm#1}}}}}

\def\refs{\vs{20}\cl{\it References}\bs\srm\bsk{12}}
\def\ref#1{\par\ni\parskip=3pt\hangindent=1.66pc\hangafter=1\hskip0.5em(#1)\sk}
\def\r#1{\par\ni\parskip=3pt\hangindent=1.66pc\hangafter=1\hskip0.0em(#1)\sk}

\def\figsnp{\vfe\topinsert\vskip0.5truein\endinsert\brm\cl{    }\vs{20}}

\def\fig#1{\srm\bsk{14}\vs{14}\cl{F{\ssrm IG}. #1}\vs2}
\def\aa#1{{\sit A\&A},\k{\sbf #1},\k}     
\def\aas#1{{\sit A\&AS},\k{\sbf #1},\k}   
\def\aj#1{{\sit AJ},\k{\sbf #1},\k}

\def\apj#1{{\sit ApJ},\k{\sbf #1},\k}

\def\mn#1{{\sit MNRAS},\k{\sbf #1},\k}

\def\obs#1{{\sit The Observatory},\k{\sbf #1},\k}
\def\pasp#1{{\sit PASP},\k{\sbf #1},\k}
\def\pdao#1{{\sit PDAO},\k{\sbf #1},\k}

\def\HD{\hbox{{\it H\hs{-0.4}D\/}}}          \def\HDs{\HD\hs3}

\def\n2000{I. Ridpath (ed.), {\sit Norton's Star Atlas and Reference Handbook},
    editions 18 \& 19 (Longman, Harlow), 1989 \& 1998}
\def\n18-20{I. Ridpath (ed.), {\sit Norton's Star Atlas and Reference 
    Handbook}, edns.~18 \& 19 (Longman, Harlow), 1989 \& 1998, Map~1;\break
    20th~edn.~(Pi, New York), 2004, p.~144.}

\def\tycho2{[Announced by] E. H\char'34g \setal \aa{355} L27, 2000.}

\def\hp{{\it Hipparcos\/}}       \def\hps{{\it Hipparcos\/}\hs4}

\def\8cat#1{A. H. Batten, J. M. Fletcher \& D. G. MacCarthy, {\sit Eighth 
    Catalogue  of the Orbital Elements of Spectroscopic Binary Systems}\break
    (\pdao{17} #1, 1989).} 
\def\8cat{A. H. Batten, J. M. Fletcher \& D. G. MacCarthy, {\sit Eighth 
    Catalogue  of the Orbital Elements of Spectroscopic Binary Systems}\break
    (\pdao{17} 1989).} 

\def\coll135#1{, in H. A. McAlister \& W. I. Hartkopf (eds.), {\sit 
    Complementary Approaches to Double and Multiple Star Research} ({\sit
    IAU Colloquium}, no.~135) ({\sit ASP Conference Series}, no.~32) (ASP,
    San Francisco), 1992, p.\hs3#1.}  
\def\adams35{W. S. Adams \setal \apj{81} 187, 1935.}

\def\f#1{The observed radial \velocities of #1 plotted as a function of phase,
    with the \velocity curve corresponding to the adopted orbital elements 
    drawn through them.\hs7}
 
\def\Herschel{I\hs{-2.5}\lower.3ex\hbox{{\smmath\char'76}\hs{-5.2}\lower.3ex
    \hbox{$_{\circ}$}}\hs{-2.3}I}
\def\H2{I\hs{-1.5}\raise0.3ex\hbox{-\hs{-2.0}-}\hs{-2.0}I\hs{-5.5}\lower.05ex
    \hbox{{\smtt\char'174}\hs{-3.25}\lower.42ex\hbox{$_{\circ}$}}\hs4}

\def\PLO16#1{W. W. Campbell \& J. H. Moore, {\sit Publ.~Lick Obs.}, {\sbf 16}, 
    #1, 1928.}

\def\Fig21{A. Baranne, M. Mayor \& J.-L. Poncet, {\sit Vistas Astr.},
    {\sbf 23}, 279, 1979.  (See Fig.~21 on p.~313.)}
\def\coll88{, in A. G. D. Philip \& D. W. Latham (eds.), {\sit Stellar
       Radial Velocities} ({\sit IAU Coll.}, no.~88) (Davis, Schenectady), 
       1985, p.~}
\def\coll170#1{ in {\sit Precise Stellar Radial Velocities,} eds.~J. B. 
    Hearnshaw \& C. D. Scarfe ({\sit IAU Coll.}, no.~170) ({\sit ASP 
    Conf.~Series}, {\sbf 185}) (ASP, San Francisco), 1999, p.~#1}

\def\hut{P. Hut, \aa{99} 126, 1981.}

\def\gren2800{[Announced by] S. Grenier \setal \aas{137} 451, 1999.}

\def\14000{[Announced by] B. Nordstr\"om \setal \aa{418} 989, 2004.}

\def\title#1{\rm\topinsert\vskip0pt\endinsert
\cl{SPECTROSCOPIC BINARY ORBITS}\vs2 \cl{FROM PHOTOELECTRIC RADIAL VELOCITIES} 
\vs8\cl{PAPER #1}\vs9\it\cl{By R.F.~Griffin}\vs{-2}\cl{Cambridge Observatories}}

\bsk{16}  \def\gD{$\gamma$~Dorad\^us}  \def\gDs{\gD\hs3}
\def\gd{$\gamma$~Dor}  \def\gds{\gd\hs3}
\cl{SPECTROSCOPIC BINARY ORBITS} \cl{FROM PHOTOELECTRIC RADIAL VELOCITIES} 
\vs3\cl{PAPER 191: HD 17310, HD 70645 {\srm AND} HD 80731}\it\bs
\settabs\+\hskip 1.3 true cm & \hskip 6.0 true cm & 2.0 true cm & \hskip 6.0 
 true cm & \hskip 1.0 true cm & \cr\it
\+ \hah By R. F. Griffin \hah and \hah H. M. J. Boffin\hah \hcr
\+ \hah Cambridge Observatories \hah \hah European Southern Observatory \hah \hcr

\vs{12}\brm\bsk{19} The three objects have been identified as members of the
recently recognized class of {\biggk\char'15}~Dorad\^us stars, which exhibit
multi-periodic photometric variations that are thought to arise from
non-radial pulsation.  The particular objects treated here also prove to be
spectroscopic binaries, for which we provide reliable orbits.  The radial
velocities exhibit unusually large residuals, in which some of the
photometric periodicities can be traced.  Some of the same periodicities are
also demonstrated by the observed variations in the line profiles, which are
quantified here simply in terms of the line-widths.

\vs{10}\rm\bsk{17}\section{Introduction}

The characters of a few stars that showed small photometric variations with
multiple periodicities of the order of one day --- longer than typical
$\delta$~Scuti variations --- gradually became apparent in the late years of
the last century.  As recently as 1999 Kaye \etal1 defined a new class of
variables, having \gDs as the type star, to accommodate such objects, whose
photometric instability has been attributed to high-order gravity-mode
pulsations, in which the motions are mainly tangential (rather than radial,
as in the case of $\delta$~Scuti pulsations, which have typical periods of
the order of 0.1 day).  The periods are in the range 0.3--3 days; the
pulsations are likely to affect the line profiles, more particularly in the
wings of the lines, but since they are largely non-radial and slower than in
the $\delta$~Scuti case their effects on stellar radial velocities are
likely to be more muted.

In the same year as the new class was recognized, Handler$^2$ presented a
list of membership candidates, identified from a comprehensive search of the
\hps `epoch photometry'; the list contained 70 entries, of which 46 were
considered to be `prime candidates'.  Many, but not all, of those that have
been investigated have proved to be spectroscopic binaries.  Paper 187$^3$
in this series gave orbits for two of them; in an introductory section it
referred to the observational history (salient parts of which were published
in this {\it Magazine\/}), of \gds itself, and to the recognition$^{4,5}$ in
9 Aur, a non-binary member of the class, of a sub-set of the photometric
periods in the star's radial velocities and line-profile variations.  One of
the present authors was also responsible for the radial-velocity
measurements that led to a double-lined orbit$^6$ for HD~221866 and the
tentative identification of the secondary component in that system as a \gds
star.

The observational histories of the three stars that form the subject of this
paper, all of which are of \HDs type F0, run extremely parallel with one
another, because (apart from some {\it uvby\/}H$\beta$ photometry, not
referenced here) it is only since their appearance in Handler's list$^2$
of `prime \gds candidates' that any interest has been taken in them.  Even
since then, there have been only three papers --- by Martin, Bossi \&
Zerbi$^7$ in 2003, Mathias \etal8 in 2004, and Henry, Fekel \& Henry$^9$ in
2005 --- that have shed much light on them; all three stars feature in the
last one$^9$, but the other two do not include HD~17310.  

\bsk{16.3}
The latter star, however, does feature in a few papers that are not
concerned with the others.  It~was among a large number of stars observed
for radial velocity by Grenier \etal{10}, who found a mean of \hbox{$-$21.4
\PN 3.6} \kmss from three \hbox{80-\amm} spectrograms, the uncertainties of
the results being such that the plate-to-plate discrepancies did not suggest
real variability; they classified the spectrum as F2\hs2IV--V.  Then Koen \&
Eyer$^{11}$ identified periods\*{In this paper we usually refer to periods
rather than frequencies, inverting where necessary the numbers quoted from
other authors.\vs{-6}} in the \hps photometry; and Nordstr\"om \etal{12}
gave basic data about the star in their survey of F and G dwarfs, but they
did not measure any radial velocities for it.  HD~70645 is mentioned in one
`extra' paper, in which Topka \etal{13} remarked that it is within a field
surveyed by the {\it Einstein\/} X-ray satellite, but no X-rays from it were
detected.  HD~80731 was observed twice by Moore \& Paddock$^{14}$ long ago
with the Lick 36-inch refractor and a prismatic spectrograph giving 75~\ahg
; they listed a spectral type of F0\hs2V and a mean radial velocity of +4
\kmss with a `probable error' of 1.7 \kms , so the binary nature of the
object was not discovered.

All three stars feature in the \hps survey.  For HD~17310, the \hps
catalogue preferred to give ground-based values of $V$ and \b-v\hs{-3},
7\mag 76 and 0\mag 378, respectively, to its own measurements, but we have
not been able to discover whence it obtained them.  The $V$ magnitude was
flagged as variable, but no period was found; Handler$^2$ and Koen \&
Eyer$^{11}$, however, thought that they saw a period of 2.0296 days in the
\hps photometry.  The parallax, \hbox{0\sec 00914 \PN 0\sec 00114,} leads to
a distance modulus of \hbox{5\mag 2 \PN 0\mag 3} and thus to an absolute
magnitude of about +2\mag 6.  HD~70645 is attributed a \hp -based mean $V$
magnitude of 8.12, appearing to vary with a period (also noted by Handler)
of 0.82488 days, and a ground-based \b-v of 0\mag 344.  The parallax of
\hbox{0\sec 00755 \PN 0\sec 00087} yields a distance modulus of {5\mag 6 \PN
0\mag 3}, implying \hbox{$M_V \sim$ 2\mag 5.}  HD~80731 is listed with
ground-based magnitudes of \hbox{$V$ = 8\mag 46,} \hbox{\b-v = 0\mag 345,}
the former variable in a period of 1.11556 days (again confirmed by
Handler), and with a parallax of \hbox{0\sec 00677 \PN 0\sec 00103,} leading
to \hbox{$m-M = 5$\mag 85 \PN 0\mag 3} and thus to $M_V \sim$ 2\mag 6.  The
three stars are all seen to have just the colour indices and absolute
magnitudes that would be expected for early-F dwarfs.

We next give a brief synopsis of what the three post-Handler papers$^{7-9}$
have discovered about the stars.  Martin \etal7 obtained fresh photometric
data on eight stars, including HD~70645 and HD~80731, on the five useable
nights during a single 14-night observing run on the 90-cm telescope in the
Sierra Nevada; the number of observations per star was only about 50.  They
also re-investigated the \hps photometry with a computer program that they
considered to be superior to other people's.  For HD~70645, Martin \ea\*{It
may save other readers of Martin \smea 's paper\up7 some time and thought if
we mention that where the caption of the all-important Table~5 refers to the
`first column' it means columns 1--6; the `second column' is really columns
7 and 8, and the `third column' is columns 9--11.  The expression
`semi-column' means a colon.} did not confirm the 0.825-day period that both
\hps and Handler thought to exist in the \hps photometry, but instead found 
periods of 0.792 and 1.297 days; their own data indicated 1.14 and 0.690
days.  For HD~80731 they confirmed the \hp /Handler periodicity of 1.1156
days in the \hps photometry, and found an additional one of 0.745 days.
Their own measures were considered to indicate possible periods of 7.00,
2.23, and 1.401 days, but it is difficult to believe that periods of such
lengths could be reliably identified on the basis of only five nights' data.

Mathias \etal8 obtained repeated spectra of a number of \gds candidates,
including HD~70645 and HD~80731, with the {\it Aurelie\/}$^{15}$ spectrograph
on the Haute-Provence 1.52-m telescope, mainly to look for line-profile
variations that would corroborate the \gds natures of the stars concerned.
They concentrated attention on just two ionic lines in the blue part of the
spectrum.  Over a total interval of a little more than a year they obtained
10 spectra of HD 70645 and 11 of HD~80731, finding line-profile variations
to be ``evident'' in both cases and discovering the binary natures of both
stars.  They derived orbits for them both; that of HD~70645 is in principle
correct, but that of HD~80731 is completely mistaken, having a period of
\hbox{13.572 \PN 0.011} days, whereas we shall show below that the true
value is 10.674 days.  Unfortunately they did not publish their radial
velocities, so we cannot take them into account in our own orbital
solutions.  We~note that we have already found$^3$ in Mathias \ea 's
paper$^8$ another mistaken period, that of HD~100215; without the data in
front of us we cannot see exactly how the errors arose, but in all cases the
number of velocities was very small upon which to base orbits.  Mean \vsinis
values of 11 and 13 \kmss were listed for HD~70645 and HD~80731,
respectively.

Henry, Fekel \& Henry$^9$ observed, in the course of a year's campaign with
an automated 0.4-m telescope at the Fairborn Observatory, all three of the
stars discussed in the present paper.  They obtained much more satisfactory
photometric coverage than either \hps (very bad temporal distribution of the
data points, but of course the photometry was only a by-product of the
principal objective) or Martin \etal7 (very small data set).  For HD~70645
and HD~80731 they had a total of more than 400 measurements in both $B$
and~$V$, obtained on about 250 nights in one season, so the derived
periodicities ought to be entirely reliable.  Martin \ea 's rejection of the
\hp - and Handler-derived period of 0.825 days for HD~70645 from the \hps
photometry in favour of 0.792 days was corroborated, and the Martin primary
period of 1.14 days was confirmed.  The Henry \eas periods for HD~70645, in
descending order of amplitude, were 1.1032, 0.7929, 0.8593, 1.2405, and
1.1461 days, all of them found in both the $V$ and the $B$ data and all
having uncertainties typically of 2 in the fourth decimal place.  In the
case of HD~80731, Henry \eas confirmed the 1.1156-day period that all of the
previous investigations had found in the \hps photmetry, but not the periods
proposed by Martin \eas from their own data.  The Henry \eas periods, in
order, were 1.1159, 1.2783, 1.5154, and 0.7623 days.

Henry \etal9 also obtained spectroscopy in the red ($\lambda \sim$ 6400 \AA
) with the Kitt Peak coud\'e-feed system.  They classified the stars,
finding both HD~70645 and HD~80731 to be of type F1 (and they knew them to
be main-sequence objects from the \hps parallaxes), and they gave \vsinis
values of 11 and 14 \kms , respectively.  They measured (and tabulated, so
we are able to utilize them) three radial velocities for HD~70645 and seven
for HD~80731.  They commented that two of their velocities of HD~70645 are
consonant with Mathias \ea 's orbit, but that the third showed a residual of
about 10 \kms , so the orbit must require some revision; and they saw that
their velocities of HD~80731 were {\it not\/} consistent with the Mathias
\eas orbit.

We have left till last our reference to the work of Henry \etal9 on the
third of the stars discussed in the present paper, HD~17310, simply because
their paper is the only one that deals with it.  The paper$^9$ lists as many
as eleven radial velocities, with a range of over 40 \kms , and gives a
spectral type of F2 and a projected rotational velocity of 10 \kms .  Three
photometric periods were established from more than 200 measurements in both
$V$ and $B$; they were 2.138, 1.825, and 2.452 days, with uncertainties near
0.001.  The 2.0296-day period derived from the \hps photometry by both
Handler$^2$ and Koen \& Eyer$^{11}$ was not confirmed.

Henry \eas performed a period search on their radial velocities of HD~17310
in an effort to identify the orbital period.  The observational `window
function' was far from ideal, because the measurements were made during just
four observing runs, in which the star was observed on one night and on
three, two, and five consecutive nights, respectively.  The best period
formally was 0.9653 days, but the authors$^9$ did not trust it.
``Instead,'' (they said) ``we prefer periods in the 20--30 day range, the
best of which is 27.793 days.''  We are able to commend both their instinct
and their conclusion, as we shall show below that that period is correct.
Their preferred period is a 1-day$^{-1}$ alias of the short one; expressed
as frequencies, they are 0.03598 and 1.03595 day$^{-1}$, respectively.

\bsk{16.5}
\section{New radial velocities and orbits}

The paper$^9$ by Henry, Fekel \& Henry, which is entitled {\it 11 New
$\gamma$ Doradus Stars}, was published in mid-2005 and caught the attention
of one of the present authors, who found particular interest in a column in
Table~1 where the \vsinis values were given for the 11 stars.  Three of the
values --- those assigned to the stars that we are discussing now --- were
from 10 to 14 \kms , whereas the others ranged from 38 to 150 \kms .  Stars
that rotate rapidly are difficult or impossible to measure for radial
velocity with the \cors at the Cambridge 36-inch telescope, but when a short
investigation of the literature had revealed that the three \gds stars that
rotated slowly lacked reliable orbits those objects were placed on the \cors
observing programme.

HD~17310 is very unfavourably placed, at a declination of nearly $-$7\degs
in Eridanus, on the border by Cetus, about 3\degs north-preceding the
fourth-magnitude star $\eta$ Eri.  Strictly speaking it ought not to be
observed with the Cambridge telescope, whose coud\'e beam is increasingly
vignetted by the telescope structure below $-$5\degs declination, but the
observer persuaded himself that the vignetting at $-$7\degs was not so great
as to be likely to produce errors as bad as those that could be expected
from other sources.  Observations were necessarily confined to the vicinity
of the meridian, so the observing season was short and yielded only 11
measurements.

HD~70645 and HD~80731, in contrast, are at high declinations (68\degs and
62\degs respectively), quite close together in the north-preceding corner of
Ursa Major.  Preliminary orbits were established very quickly, and for a
time thereafter the radial velocities were measured only where they would
serve to fill gaps in the phase distribution.  Later, when it became
apparent that the residuals from the orbits might yield (or at least
exhibit) some of the pulsational periodicities, a measurement was made on
each fine night regardless of orbital phasing.  The two stars, though easily
circumpolar as seen from Cambridge, cannot be observed more than about 6
hours from upper culmination because structures associated with the
northward-going coud\'e focus (and indeed the \cors instrument itself)
occupy the northern part of the dome of the 36-inch reflector.  The total
numbers of new radial velocities are 44 for HD~70645 and 54 for HD 80731.

Although the data are (unusually, for this series of papers) confined to a
single observing season, by reason of their continuity and compact
distribution in time they lend themselves tolerably well to the
investigation of short periods superimposed upon the orbital variation.  The
likelihood that pulsational instabilities would be traceable in the radial
velocities is indicated by substantial variations in the profiles of the
cross-correlation dips from which the velocities are determined.  That is
illustrated by Fig.~1, which compares the dips given by HD~80731 on
different occasions.  The $S/N$ ratios achievable for radial-velocity traces
of the stars concerned are not usually adequate to delineate with confidence
any real asymmetries that may be present, but the overall widths of the line
profiles certainly change from one occasion to another.  The widths are
characterized numerically here as if they were projected rotational
velocities, \vsini , but our use of that expression is to be regarded merely
as a name for the line-width parameter that is routinely calculated for each
radial-velocity trace by the \cors reduction software {\it as if\/} the
broadening of the spectral lines, beyond the minimum width given by other
stars, were due simply to rotation of the stars as solid bodies.  The values
are quantized in \hbox{\frac1/2-\kmss} steps, owing to the manner in which
they are calculated$^{16}$.

Straightforward orbital solutions of the radial-velocity data give orbits
that are quite satisfactory but are characterized by unusually large
residuals, of the order of 1.5 \kmss or so --- two if not three times as
large as might be expected from the character of the data.  We show below
that pulsational periods can be traced in the residuals.  We can try to
model the residuals by sine waves, by regarding the residuals as radial
velocities in their own right and solving them with a program that derives
circular orbits from such data.  If there were only one short period, it
would perhaps be appropriate to treat the raw radial-velocity measurements
with the orbit program that solves simultaneously the outer and inner orbits
of single-lined triple systems.  That, however, manages to improve slightly
the fit to the short-period, low-amplitude `inner orbit' by making slight
changes to the `outer' --- in this case, the only true --- orbit.  Not only
would such changes not be likely to suit more than one periodicity among the
pulsational `orbits' but, as the number of data points increases, the scope
for adjusting the true orbit to accommodate residuals arising from pulsation
decreases until in the limit of an indefinitely large number of data it
would vanish altogether.  We conclude that the proper procedure is first to
derive the actual orbit by a straightforward application of the single-line
orbit-solving program, and then to use the resulting set of residuals as the
dataset to be investigated for evidence of pulsational periodicities.  In
the sections below, we discuss the stars out of conventional right-ascension
sequence in order to treat the two comparatively well-observed ones first.

It was only as our observations accumulated, and we realized that the
velocity residuals were not random but exhibited one or more periods
associated with the \gds pulsations, that concern arose as to whether the
quality of the data would be adequate to support an analysis of pulsations
whose amplitudes would be very much smaller than those of the orbits that we
initially set out to determine.  Most of the later observations, therefore,
were integrated to more generous levels, usually $>$ 10\hs2000 counts per
bin, than most of the earlier ones, which were nearer 5000.  In analyzing
the final datasets, we experimented with flagging them ($a$) by temporal
halves, and alternatively ($b$) by the count levels.  There proved not to be
significant differences between any of the divisions thus made:\hs3 the
conclusion to be drawn from the exercise seems to be that observational
error is {\it either\/} not the principal contributor to the velocity
residuals (the analysis of which is therefore valid), {\it or\/} is not
significantly reduced by approximately doubling the integrations (which is
difficult to believe).  It also appears, therefore, that our concern over
the data quality was misplaced, and that the extra time spent on many of the
later integrations may largely have been wasted!

\section{HD 70645}

The Cambridge radial-velocity measurements began in 2005 November, soon
after the star was first observable on the dawn meridian, and continued
until it was beyond reach in the north-west at dusk at the beginning of the
following June.  Forty-four observations were made of it; they are listed in
Table~I, together with the three velocities published by Henry \etal9 and
the phases and residuals obtained from a solution performed as for a normal
single-lined binary star.  All the velocities were given equal weight.
Initially the Cambridge measurements were solved alone, and gave the period
with a standard error of 0.0022 days.  Then the published data were brought
in to the solution, producing negligible changes to the elements but (by
increasing the time base from 200 to 1200 days) reducing the standard error
of the period to 0.0008 days, a worthwhile improvement.  The solution is
illustrated in Fig.~2 and the orbital elements are set out in the informal
table below.  

\bs\settabs\+\hskip 1.5 true cm & \hskip 0.7 true cm &
\hskip 7.0 true cm & \hskip 1.5 true cm & \cr
\+ & $P$ & = \sk 8.4402 \PM 0.0008 days &
$(T)_{10}$ & = \sk MJD 53764.26 \PM 0.21 \cr
\+ & $\gamma$ & = \sk +14.53 \PM 0.28 \kms &
\asini & = \sk 3.82 \PM 0.05 Gm \cr
\+ & $K$ & = \sk 33.1 \PM 0.4 \kms &
$f(m)$ & = \sk 0.0314 \PM 0.0011 \msun \cr
\+ & $e$ & = \sk 0.077 \PM 0.012 \cr
\+ & $\omega$ & = \sk 66 \PM 9 degrees &
R.m.s. residual \sk =\sk 1.72 \kms \cr\bs

\ni Unfortunately the potential pulsational periods cannot be determined
accurately enough for the cycle count back to the published observations to
be secure, so from this point onwards the investigation is limited to the
Cambridge measures.  A column has been added to Table~I to give the apparent
\vsinis value determined individually from each observation.

We have treated the velocity residuals, with their corresponding
observational epochs, as an autonomous dataset, to be examined for
pulsational periodicities as explained at the end of the section above.
Equally, we regarded the line-widths as constituting a parallel dataset
meriting an analogous examination.  Rather than choosing whether to test for
the presence of periods already proposed by others, or instead to make an
independent search for periodicities, we decided to adopt first the one
strategy and then the other.  It could be argued that the photometric
periods found by Henry \eas come from such a rich database that their
validity in the magnitude data is practically guaranteed, so all we need to
do is to test for their presence in the radial velocities and the
line-widths; but it would be a pity to overlook other periods, that might be
more conspicuous in radial velocities or in line-widths than in brightness,
simply by neglect of an unprejudiced search of our own data.  At the
same time we need to be careful not to fall into the error, of which we have
sometimes$^{18,19}$ suspected others, of placing too much reliance on short
periods that may be mathematically present in sparse data strings.

Our procedure for assessing the significance of possible periods, whether
taken from the literature or found by ourselves, was as follows.  We set up
the datasets as if for single-lined solutions of circular orbits, with the
period to be tested, a nominal amplitude of 1 \kms , and with the epoch
set at that of the largest positive (residual) velocity or the largest
line-width.  If the solution then ran and converged, that result was taken
as qualitative evidence in favour of the existence of the relevant period in
the data.  If it did not, we ran a `plot-only solution', in which we did not
ask the computer to improve the elements that we had supplied but simply to
plot the solution as it stood; that would enable us to see from the plot
whether there appeared to be a significant variation with the relevant
period but not with the phasing implied by our inevitably crude guess.  Any
apparent variation could be followed up by re-running an optimized solution
with an appropriately adjusted initial epoch.  Where no evidence of a
systematic phase-related variation could be seen in a `plot-only solution'
and the computer could not be persuaded to pull in to any solution and
improve on it, we concluded that no significant periodicity existed.  We
feel quite secure in doing that, since (as we proceed to show) several of
the results where there {\it did\/} appear to be {\it some\/} evidence of
phase dependence, and where the computer {\it did\/} grasp the solution and
improve on it, have turned out to be without statistical significance.

The method that we have selected for quantifying significance is to compare,
in the light of the $F$ test, the sums of the squares of the residuals from
the solutions obtained with and without the prospective period.  We consider
first the radial-velocity case.  The `without' sum is always the same, being
just the sum of the squares of the residuals that are given in Table~I for
the 44 Cambridge observations and that form the data set that we are testing
for periodicities.  That sum is 132.24 (\kms )$^2$.  In solving those data
for a circular `orbit' we attach optimal values to four independent
variables, \viz period, epoch, amplitude, and \g .  (Although, from the
manner in which they were obtained, the velocities that constitute the
dataset must have a weighted mean of zero, it does not follow that the \gs
of an optimized pulsational `orbit' derived from them will be exactly zero.)
Especially if the imposed or resulting period is a significant one, the sum
of squares of the new set of residuals will be reduced.  The amount of the
reduction will be assignable to the four degrees of freedom represented by
the four fitted variables, while the remaining sum is to be laid at the door
of the other 40 degrees of freedom; the significance of the reduction is
found by taking the ratio of the mean-square per degree of freedom between
the four and the 40 and comparing it with tabular values of $F_{4,40}$ for
various levels of significance.

We clarify the procedure by an initial `worked example' based on the most
significant pulsational period that we have found for HD~70645, \viz the
0.792-day one found first by Martin \etal7 and corroborated by Henry \etal9.
The computer pulls into a solution with a period of \hbox{0.7919 \PN 0.0003}
days and an amplitude of \hbox{1.62 \PN 0.31 \kms } (which at $>$5$\sigma$
looks promising) and yields a sum of squares of 78.81 (\kms )$^2$ for the 44
residuals, which represent the 40 degrees of freedom left after we used four
in fitting the four variables.  Thus the 40 degrees cost \hbox{78.81/40 =
1.97} (\kms )$^2$ each.  The four degrees represented by the variables cost
\hbox{(132.24 $-$ 78.81),} or 53.43 (\kms )$^2$, \ie 13.36 per degree, so we
obtain \hbox{$F_{4,40}$ = 13.36/1.97} = 6.80.  The tabular values$^{17}$ of
$F_{4,40}$ for various degrees of significance are as follows: 10\% 2.09,
5\% 2.61, 2\frac1/2\% 3.13, 1\% 3.83, 0.5\% 4.37, 0.1\% 5.70.  Our value is
therefore comfortably beyond even the 0.1\% value of $F$, and it follows
that the period is to all intents and purposes certainly present in the
data.  The plot of the fitted sine-wave and the velocities to which it is
fitted (the residuals from the orbit derived above and plotted in Fig.~2) is
shown in Fig.~3.

We have been through all of the periods found by Henry \eas in the same way,
and present the results very succinctly in Table~II.  The successive lines
of the Table give the successive quantities specified in our illustration of
the procedure above, as follows:\ms

\def\b{\hm1} \def\bb{\hm{1.5}} \def\bbb{\hm2}\def\c{\hm{0.5}} \def\d{\hm{4.27}}

(a)\b Period (days) given by Henry \etal9;

(b)\b Period (days) found in our velocities, with its standard error in
units of the last decimal place in brackets;

(c)\b Amplitude (\kms ) found in our velocities, with its standard error
similarly;

(d)\b Sum of squares of the velocities ((\kms )$^2$) before the period is
fitted;

(e)\b Sum of squares after the period is fitted, followed by that quantity
divided by 40 (the number of degrees of freedom that it represents), so the
second number is the mean square per degree of freedom;

(f)\b (d) minus (e), the remaining portion of the sum of squares,
attributable to the four degrees of freedom represented by the fitted
period, and the same quantity divided by four to give the mean square per
degree; 

(g)\b the ratio of the mean squares in (e) and (f) immediately above,
=~$F_{4,40}$; and finally

(h)\b the significance of that $F$ ratio (n.~s. = `not significant').

\ni The ensuing lines (b1) to (h1) will be explained shortly.\ms

\bs\cl{T{\srm ABLE} II}
\it\cl{Significances of periods in the radial- and rotational-velocity data
on HD 70645}\bs\rm\tabskip4em
\vbox{{\srm\bsk{10}
\halign{\hm2#\hf&#\hf&#\hf&#\hf&#\hf\cr
(a)&\b0.7929&\b0.8593&\b1.1032&\b1.2405\cr
(b)&\b0.7919\b(3)&\b0.8606\b(5)&\b1.1049\b(8)&\b1.2428\c(16)\cr
(c)&\b1.62\bb(31)&\b1.35\bb(34)&\b1.17\bb(37)&\b0.93\bb(37)\cr
(d)&132.24&132.24&132.24&132.24\cr
(e)&\c78.81\bb1.97&\c95.03\bb2.38&105.99\bb2.65&113.52\bb2.84\cr
(f)&\c53.43\b13.36&\c37.21\bb9.30&\c26.25\bb6.56&\c18.72\bb4.68\cr
(g)&\d6.80&\d3.91&\d2.48&\d1.64\cr
(h)&\hm{0.25}0.1\%&\b1\%&\hm{0.75}10\%&\b n.~s.\cr
\noalign{\vs9}
(b1)&\b0.7927\b(5)&\b0.8594\b(9)&\b1.1018\c(10)&\b1.2475\c(16)\cr
(c1)&\b1.26\bb(35)&\b0.84\bb(38)&\b0.99\bb(38)&\b0.9\b\bb(4)\cr
(d1)&134.94&134.94&134.94&134.94\cr
(e1)&100.90\bb2.52&120.48\bb3.01&114.37\bb2.86&120.26\bb3.01\cr
(f1)&\c34.04\bb8.51&\c14.46\bb3.61&\c20.57\bb5.14&\c14.68\bb3.67\cr
(g1)&\d3.38&\d1.20&\d1.79&\d1.22\cr
(h1)&2\smfrac1/2\%&\b n.~s.&\b n.~s.&\b n.~s.\cr}}}

\bs\bsk{17.5}
It is to be noticed that the order of the periods listed in Table~II with
progressively decreasing significance is not the same as the order of the
photometric amplitudes found by Henry \ea , which is 2, 3, 1, 4 for the
succesive columns in our Table; the uncertainties in the photometric
amplitudes, however, are large enough in relation to the amplitudes
themselves to mean that the ordering of the photometric periods is not
really determinate.  Henry \ea 's fifth period is omitted from our Table
because it did not produce a plot that looked at all significant, and the
sum of squares fell only to about 125, so there is no evidence for its
presence~at~all.  

An analysis can be made of the significance of the various periods in the
\vsinis data in exactly the same fashion as for the radial velocities.  
The mean \vsinis is 12.16~\kms , and the r.m.s.~spread of the individual
values is 1.75~\kms , the mean square being therefore (1.75)$^2$ or 3.07, and
the sum of squares 44 times that, 134.94 (\kms)$^2$ (coincidentally very
close to that of the radial velocities); that total is apportioned between
the four degrees of freedom represented by the fitted sine-wave and the
remaining 40 degrees, exactly as in the case of the radial velocities, so we
have simply added to Table~II another set of lines corresponding precisely
with the first set, labelled (b1) to (h1) and pertaining to the \vsinis
data.

We point out that potential periodicities may lack statistical significance
but nevertheless be present in the data.  That may be suggested in some
cases by the simple fact that the attempt to compute a solution with a given
period does actually produce convergence, and does so at a period that is
close to the suggested one.  If~the process is initiated with an arbitrary
period, it tends either to diverge or to pull into a period that is not
plausibly similar to the one under trial.

The final results of our analysis, seen in lines (h) and (h1), are that
three of the Henry \eas photometric periods are traceable in the radial
velocities, with diminishing degrees of significance, but only one is
significant in the rotational velocities.  It is the same period that is
much the most signficant one in the radial velocities that is also traceable
in the rotational ones.  As a general comment on the results of Table~II and
the analogous tables to follow for the other two stars, we remark that the
significances that are found from the $F$ test are smaller than might be
anticipated from a comparison of the amplitudes in lines (c) and (c1)
with their respective standard errors.  Although we cannot offer any
mathematical reason for the apparent discrepancy in significances, we
understand that it is well known to period-search experts that ratios less
than 4 for $K$~to~$\sigma$($K$) are not usually significant, notwithstanding
that in a `normal distribution' a significance of 1\% is reached at a ratio
of 2.58.

\section{HD 80731}

Just as in the case of HD 70645, radial-velocity measurements with the
Cambridge \cors began in 2005 November and continued until the observing
season closed in the following June, and (again like HD~70645) it was
observed with increased assiduity towards the end of the season in order to
improve the chances of documenting pulsational periods.  The Cambridge
measurements number 54; they are set out in Table~III, along with the seven
velocities published by Henry \etal9 as well as with the phases and
residuals that stem from a normal single-lined orbital solution, and also
with the `\vsini ' values for each of the observations.  A~solution based on
the Cambridge observations alone gave a period of \hbox{10.678 \PN 0.004}
days; the inclusion of six out of the seven Henry \etal9 measures did not
change the elements significantly but reduced the standard error of the
period to 0.0004 days.  Among the published measurements, the third one has
such an extreme residual (9~\kms , more than twice as great as any other)
that it seems to be beyond any combination of pulsational velocities plus
normal accidental eror, so it may be suspected of some sort of qualitative
error and has therefore been omitted from the solution.  The orbit has the
\l1 elements given below and is plotted in Fig.~4.

\bs\+ & $P$ & = \sk 10.6744 \PM 0.0004 days &
$(T)_4$ & = \sk MJD 53731.19 \PM 0.05 \cr
\+ & $\gamma$ & = \sk $-$0.78 \PM 0.22 \kms &
\asini & = \sk 3.08 \PM 0.05 Gm \cr
\+ & $K$ & = \sk 22.82 \PM 0.33 \kms &
$f(m)$ & = \sk 0.0103 \PM 0.0005 \msun \cr
\+ & $e$ & = \sk 0.392 \PM 0.012 \cr
\+ & $\omega$ & = \sk 265.7 \PM 2.3 degrees &
R.m.s. residual \sk =\sk 1.57 \kms \cr\bs

To investigate the presence of pulsational periods we have followed exactly
the same procedure for HD~80731 as for HD~70645, so we can proceed
immediately to present the results, which are shown in Table~IV.  The first
three periods given in line (a) are those of Henry \etal9, and the fourth is
one of those proposed by Martin \etal7.  As in Table~II, the first section
(as far as line (h)) refers to pulsations seen in the radial-velocity
residuals, while the second section (lines (b1) -- (h1)) refers to those
seen in the rotational velocities.  The third section (lines (b2) to (h2))
gives information about two additional periods identified by ourselves in
the rotational velocities (clearly line (a) does not apply there).
Opportunity is taken to use the spare space in that section to include brief
reminders of the significance (described in full immediately before
Table~II) of the successive lines.

\bs\cl{T{\srm ABLE} IV}
\it\cl{Significances of periods in the radial- and rotational-velocity data
on HD 80731}\bs\rm
\vbox{{\srm\bsk{10}
\halign{\hm{1.5}#\hf&#\hf&#\hf&#\hf&#\hf\cr
(a)&\b1.1159&\b1.2783&\b1.5154&\b2.23\cr
(b)&\b1.1159\b(9)&&\b1.5157\c(22)&\b2.236\bb(4)\cr
(c)&\b1.00\bb(32)&&\b0.91\bb(28)&\b1.05\bb(28)\cr
(d)&137.1&&137.1&137.1\cr
(e)&113.9\bbb2.28&&112.5\bbb2.25&106.0\bbb2.12\cr
(f)&\c23.2\bbb5.80&&\c24.6\bbb6.15&\c31.1\bbb7.77\cr
(g)&\d2.55&&\d2.73&\d3.67\cr
(h)&$\sim$5\%&&\b5\%&2\smfrac1/2\%\cr
\noalign{\vs9}
(b1)&\b1.1147\b(7)&\b1.2774\c(14)&\b1.5137\c(25)\cr
(c1)&\b2.3\b\bb(5)&\b1.6\b\bb(5)&\b1.4\b\bb(6)\cr
(d1)&432.1&432.1&432.1\cr
(e1)&282.1\bbb5.64&357.9\bbb7.16&385.6\bbb7.71\cr
(f1)&150.0\bb37.05&\c74.2\bb18.55&\c46.5\bb11.62\cr
(g1)&\d6.71&\d2.59&\d1.51\cr
(h1)&\hm{0.25}0.1\%&\b5\%&\hm{0.75}n.~s.\cr
\noalign{\vs9}
(b2)&\b0.8210\b(5)&\b1.1617\b(7)&\hm{-2}\rlap{Period and (in brackets) its
standard error}\cr
(c2)&\b1.9\b\bb(4)&\b2.2\b\bb(5)&\hm{-2}\rlap{Pulsational amplitude and its
standard error}\cr
(d2)&432.1&432.1&\hm{-2}\rlap{Sum of squares, apportioned between:}\cr
(e2)&314.3\bbb6.29&328.5\bbb6.57&\hm{-1}\rlap{from 50 degrees of freedom, and per degree}\cr
(f2)&117.8\bb29.45&103.6\bb25.90&\hm{-1}\rlap{from 4 degrees (pulsation), and per degree}\cr
(g2)&\d4.69&\d3.95&\hm{-2}\rlap{{\sit F}\down{4,50} (quotient of the above 2 lines)}\cr
(h2)&\hm{0.25}0.5\%&\hm{0.75}1\%&\hm{-2}\rlap{Significance of the {\sit F\/} ratio above}\cr}}}\bs

\section{HD 17310}

Although the first Cambridge radial-velocity measurement was made in 2005
September, it was not till late November that routine measurements began,
and after the observer was absent for much of 2006 January the observing
season was practically at its close.  There are only 11 Cambridge
measurements, listed in Table~V, to add to the same number published by
Henry \etal9.  The Cambridge data are, however, better distributed in time,
and when solved by themselves for the orbital elements they yield an
unambiguous period of \hbox{27.67 \PN 0.22} days, very similar to the best
(but not uniquely determined) value of 27.793 days favoured by Henry \eas
\hs3 The precision of the Cambridge period is plenty good enough to
extrapolate back to the Henry \eas epochs without any possible error in the
cycle count, so the two sets of velocities can be solved together.

The joint solution does, however, throw up problems that were mercifully
lacking in the cases of the other two stars.  For them the data from the two
sources seemed to agree well both in zero-point and in the sizes of the
residuals from the orbits, so neither a zero-point shift nor unequal
weighting was called for.  In the present case, a straightforward solution 
with equal weights shows fairly serious disparities both in zero-point and
in residuals.  The means of the residuals from the two sources differ by
\hbox{1.62 \PN 1.16} \kms , Cambridge being more positive, while the mean
squares (the variances) are 3.81 and 10.82 (\kms )$^2$ for Cambridge and
Henry \ea , respectively, a difference of a factor of 2.85.  With six
elements fitted to 22 equally weighted observations, we could consider that
each source has 8 degrees of freedom; the factor that we have found exceeds
the 10\% point of $F_{8,8}$, which is 2.54, so we have seen fit to attribute
half-weight to the published velocities.  The weighted variances then turn
out to be 2.93 and 6.05 (\kms )$^2$, a ratio still as high as 2.06, but no
longer very significant; we would have to go a lot further to equalize them,
but we do not care to do that, particularly in the light of the apparent
quasi-equality in the cases of the other stars.  We could also worry about
the zero-points, which in the revised (weighted) solution differ by
\hbox{1.84 \PN 1.15} \kms , or \hbox{1.60\hs1$\sigma$,} for which the 
probability according to the `normal distribution' is about 11\%.  Not
wishing to tamper too much with an already minimal dataset, we decided {\it
not\/} to make any adjustment to the relative zero-points.  On the basis,
then, of no interference with the observed velocities apart from
half-weighting the published ones, we obtain the orbit that is plotted in
Fig.~5 and whose elements are:\bs

\+ & $P$ & = \sk 27.819 \PM 0.022 days &
$(T)_{-5}$ & = \sk MJD 53485.4 \PM 1.2 \cr
\+ & $\gamma$ & = \sk +25.8 \PM 0.7 \kms &
\asini & = \sk 9.0 \PM 0.4 Gm \cr
\+ & $K$ & = \sk 24.0 \PM 1.0 \kms &
$f(m)$ & = \sk 0.038 \PM 0.005 \msun \cr
\+ & $e$ & = \sk 0.19 \PM 0.04 \cr
\+ & $\omega$ & = \sk 125 \PM 15 degrees &
R.m.s. residual \sk =\sk 2.1 \kms \cr\bs

For the purposes of searching for pulsational effects, potentially of
multiple periods that are all of short periods and small amplitudes, our
data are woefully few.  The three periodicities tabulated by Henry \etal9
and the one identified in the \hps `epoch photometry' by Koen \&
Eyer$^{11}$, however, are all near 2 days (not 1 day, as in the cases of
HD~70645 and HD~80731), and that circumstance makes it more plausible to
search the whole run of 22 velocities for the relevant periods.  There are
gaps of about one year each, or 180 cycles, in the data set, and the Henry
\eas periods being tested have uncertainties between one and two thousandths 
of a day, so it looks as if the phasing error after a year is only of the
order of 0.3 days and there should be little chance of getting out of phase
by whole cycles.

We did first test the various periods against the Cambridge data alone, but
since we used up six degrees of freedom by determining the basic orbit from
the 11 observations, by the time we had fitted another `orbit' (circular, so
using 4 degrees of freedom) to the observations there could be said to be
almost nothing left --- with one more variable one could in principle fit
{\it all\/} the velocities exactly!  To do so, however, one would need to
solve the two orbits simultaneously rather than {\it seriatim\/} as in our
tests.  Having determined the orbit of the spectroscopic binary, we could
possibly regard the residuals as constituting a fresh data source,
notwithstanding that those residuals will have been reduced (\ie the
information in which we are then interested will have been diluted) by some
accommodation of the pulsational velocities by the orbital solution.  The
upshot of the investigation, in any case, was that we found nothing
significant at any of Henry \ea 's three periods, but there was a remarkably
large signal at the Koen \& Eyer period of 2.0296 days.  Indeed, a
periodicity very close to 2 days is conspicuous in the data of Table~V just
upon inspection --- the violent reversal of the signs of residuals between
alternate days gives it away --- and a period of 1.966 days, that gave an
even more dramatic signal than the Koen \& Eyer period, was noticed before
any period-search program was brought to bear.  The search program
identified a still more potent period at 0.6618 days.  The actual
statistical significances of the various periods are very doubtful owing to
the scarcity of data, and comment is withheld until we have presented an
analysis of the complete dataset including the Henry \eas velocities.

\bsk{17}
Before going on to do that, however, we may refer to the `\vsini ' data
for HD~17310, of which there are just the 11 Cambridge values.  Their mean
is \hbox{9.6 \PN 0.8} \kms , and the sum of the squares of the deviations
from that value is about 65 (\kms )$^2$.  We can state the results of our
trials quite briefly.  The three Henry \eas periods did not reduce the sum
of squares to any great extent, but the periods of 2.0296, 1.966, and 0.6618
days all produced sums of squares reduced to near 20 (\kms )$^2$.  Since the
rotational-velocity numbers were not utilized in the determination of the
binary-star orbit, the 11 values could reasonably be regarded as independent
data and therefore as possessing jointly 11 degrees of freedom, of which
four are used up in fitting any pulsational `orbit'.  So, in exact analogy
with the treatment described before the presentation of Table~II for
HD~70645, we may say that the sum of squares remaining after the derivation
of a pulsational periodicity is associated with seven degrees of freedom,
while the reduction from the original sum represents the cost of the four
degrees used in the fit.  In that case, the three periods that caused
reductions of about 45 and left sums of about 20 (\kms )$^2$ gave values of
$F_{4,7}$ of about (45/4)/(20/7), $\sim$ 4, which is nearly the 5\% point
(4.12).  Thus, in view of the fact that we were not trying to find fresh
periods but were merely making individual tests of ones that had been
proposed on the basis of quite independent data, there are grounds for
cautious optimism in thinking that those periods may be discernible in the
rotational velocities.  Conversely, the likelihood of the existence of the
periods in the rotational velocities provides some support for their
presence in the radial ones.

We next extend the investigation of periodicities in the orbital
radial-velocity residuals to the complete dataset, including the Henry \eas
velocities.  The results are presented in exactly the same way as for the
other two stars, in Table~VI below.  In the first part of the table we test
the four periods (three from Henry \eas and one from Koen \& Eyer) that have
been identified photometrically, and then give the results from the other
two periods that seemed so significant in the Cambridge radial-velocity data
and gained some support from the rotational velocities.  The figure for the
sum of squares of the `raw' orbital-velocity residuals, appearing throughout
in lines (d) and (d1), is seen to be 98.51 (\kms )$^2$.  The 1\% point of
$F_{4,18}$ that is very nearly reached by two of the periods in the first
section is 4.58.  The 0.1\% point is 7.46, so both the values in the second
section of the table are far beyond that.  Even the highest significance
that we have seen tabulated for $F_{4,18}$, 0.05\%, is `only' 8.47.  The
periods quoted in line (a1) are those found first, when the Cambridge
velocities were considered alone, and are otherwise unsupported.  The
1.968-day period is illustrated in Fig.~6.

\bs\cl{T{\srm ABLE} VI}   \def\d{\hm{3.77}}
\it\cl{Significances of periods in the radial-velocity data on HD 17310}
\bs\rm\tabskip4em
\vbox{{\srm\bsk{10}
\halign{\hm2#\hf&#\hf&#\hf&#\hf&#\hf\cr
(a)&\c2.137&\c1.825&\c2.451&\c2.0296\cr
(b)&\c2.1360\b(7)&\c1.8220\b(4)&\c2.4510\c(10)&\c2.0292\b(4)\cr
(c)&\c1.8\bb\b(7)&\c2.4\bb\b(6)&\c1.9\bb\b(7)&\c2.6\bb\b(6)\cr
(d)&98.51&98.51&98.51&98.51\cr
(e)&71.57\bb3.97&49.40\bb2.74&68.93\bb3.83&49.77\bb2.76\cr
(f)&26.94\bb6.73&49.11\b12.28&29.58\bb7.39&48.74\b12.18\cr
(g)&\d1.70&\d4.49&\d1.93&\d4.41\cr
(h)&\c n.~s.&almost 1\%&\c n.~s.&almost 1\%\cr
\noalign{\vs9}
(a1)&\c1.9650\c(26)&\c0.6617\b(2)&\hm{-2}\rlap{Period being tested and it
standard error}\cr
(b1)&\c1.9683\b(3)&\c0.66173\c(3)&\hm{-2}\rlap{Period found and its standard
error}\cr
(c1)&\c2.9\bb\b(4)&\c3.3\bb\b(4)&\hm{-2}\rlap{Pulsational amplitude and its
standard error}\cr
(d1)&98.51&98.51&\hm{-2}\rlap{Sum of squares, apportioned between:}\cr
(e1)&24.65\bb1.36&21.27\bb1.18&\hm{-1}\rlap{from 18 degrees of freedom, and per degree}\cr
(f1)&73.86\b18.46&77.24\b19.31&\hm{-1}\rlap{from 4 degrees (pulsation), and per degree}\cr
(g1)&\d13.6&\d16.3&\hm{-2}\rlap{{\sit F}\down{4,18} (quotient of the above 2 lines)}\cr
(h1)&\hm{0.25}0.1\%&\hm{0.25}0.1\%&\hm{-2}\rlap{Significance of the {\sit F\/} ratio above}\cr}}}\bs

\section{Discussion}

We have gone some way towards demonstrating, by a formal (if elementary)
statistical analysis, that certain periods, mostly already recognized in
photometric datasets that are much richer than our kinematic ones, are
present in the radial and quasi-rotational velocities that we have measured
for the three \gds stars.  There remain questions, however, as to how far
the statistical results should be trusted.

We have already indicated an inclination towards trusting them where we are
simply testing already-defined periods for their presence in our data.  In
such cases we are not searching a sparse data string for a short period, a
procedure that we {\it know\/}$^{18}$ can lead to `false positives'.  If the
test calculation immediately converges and gives a period that is, within
the joint uncertainties of itself and of the trial period, the same as the
one being tested, there are grounds for thinking that the result is secure.
Misgivings start to creep in, however, when we consider the results of
multiple periodicities identified in the same dataset.  If we look at row
(f) of Table~II, for example, we see that the sum of contributions listed as
being made by the four tested periods to the total sum of squares is more
than that whole total!  A greater excess of the individual contributions
over the whole total is seen in row (f) of Table~VI.  It could, however, be
argued that we should not include contributions from periods that have
turned out not to be significant.  If we pretend for a moment to be really
na\"\i ve operators, we could imagine ourselves trying any number of periods
at random, and most of them would yield a `solution' that was better than
nothing, in the sense that it would produce {\it some\/} reduction in the
sum of squares; but it would be nonsensical to add up all the reductions and
say that we had thereby accounted for all and more of the apparent
raggedness of our velocities.  Clearly some consideration ought to be given
to the {\it total\/} number of degrees of freedom used in fitting multiple
periods to the same data, but we are unable to suggest how to do that in a
constructive fashion.

A more extreme situation than those already mentioned is the one referred to
in the paragraph next but one before Table~VI above, where each of three
periods is apparently found to be responsible for more than two-thirds of
the total sum of squares!  Two of those periods, however, have found no
support from photometry and might be dismissed as mere idiosyncrasies in the
very small Cambridge dataset, especially as they result from doing just what
we have warned against$^{18,19}$, \viz searching a data string for periods
short in comparison with the mean interval between observations.  But in
that case why would they be overwhelmingly reinforced when the dataset in
which they were first noticed was expanded to include the published
velocities?  Row (f1) in Table~VI appears to show that each of the two `new'
periods accounts for three-quarters of the total variance!

We can see that at least part of the answer is that there is really only one
period:\hs3 the two new ones are 1-day$^{-1}$ aliases of one another
(although not within their joint formal uncertainties).  Labelling 1.9683
days as $P_1$ and 0.66173 days as $P_2$, we find the corresponding
frequencies to be \hbox{$\nu_1$ = 0.5081} and \hbox{$\nu_2$ = 1.5112}
day$^{-1}$, respectively.  Moreover, the Koen \& Eyer period of 2.0296 days
($P_3$) inverts to --- in fact it was actually given by those authors as ---
a frequency of 0.4927 day$^{-1}$ ($\nu_3$), which is seen to be very closely
the 1-day$^{-1}$ complement of $\nu_1$.  The close numerical relationships
between all three of the periods that seem to be so powerfully present in
the small dataset of HD~17310 radial velocities warns us of the likelihood
that at most one of the three periods can be real, the others being mere
mathematical artefacts.  When plotted {\it modulo\/} the three periods in
turn, the \hps `epoch photometry' seems unrelated to $P_2$, but at least to
a subjective view its phase-dependence on $P_1$ is scarcely less convincing
than that on $P_3$, which itself inspires little confidence; the Henry \eas
periods are even less visible in the \hps photometry.

Clearly we cannot claim to have the last word, let alone the greatest
wisdom, on these matters, which would become clearer, even in the absence of
fresh insight or inspiration, if we could bring to bear a much greater
quantity of data.  What we {\it can\/} claim in this paper is to have
established the spectroscopic orbits of the three stars; in descending order
of certainty we believe also that we can trust the demonstration of two of
the already-known photometric periods in the radial velocities of HD~70645,
and probably also in HD~80731, and we think that we have traced the dominant
radial-velocity period in each of those stars in the line-width parameter
too.  It furthermore seems likely, from our very parsimonious data, that at
any rate two of the four photometric periods that have been identified in
HD~17310 are present in the radial velocities; one of them appears also to
be present in the line-widths.  Those widths, as well as the radial
velocities, are also represented extraordinarily well by either of two
periods that are aliases of one another and of one of the photometric ones,
but we are not able to adjudicate on the reality of those periods.

As far as our observations (or those of Mathias \etal8) are concerned, all
three of the systems with which we are concerned are single-lined.  We have
not noticed secondary dips in any of the radial-velcity traces, although we
regret having omitted to make a specific search for them by taking long
integrations at the appropriate velocity ranges near the nodes of any of the
orbits.  We can say only that the secondaries are probably at least two
magnitudes fainter than the primaries.  None of the mass functions is
particularly large:\hs3 those of HD~70645 and HD~17310 are both between 0.03
and 0.04~\msun , while that of HD~80731 is only 0.01.  For an early-F star
whose own mass may be estimated at 1.6 \msun , a mass function of
0.04~\msuns requires the secondary to have a minimum mass of about 0.55
\msun , corresponding to a spectral type not much earlier than M0\hs2V and
an absolute magnitude about six magnitudes fainter than that of the primary.
If~the orbital inclination is far from 90\dego , however, the secondary may
be more massive than the minimum value, to any extent, though statistically
inclinations are high.

In their study of 59 potential \gds stars whose candidatures were mostly
based on the \hps photometry, Mathias \etal8 were able, in a very few cases,
to identify the main \hps frequency in their radial-velocity curves.  For
those cases, they deduced ratios between the radial-velocity and photometric
amplitudes in the range 35 to 96 \kmss per magnitude.  We can perform the
same exercise for the three stars whose radial velocities we have analyzed,
all of which have periods in common with the photometric ones found by Henry
\etal9.  For HD~70645, we find ratios of radial velocities to $B$- and
$V$-band signals between 56 and 81, and 74 and 110 \kmss per magnitude,
respectively. The corresponding numbers are 33 to 95 and 42 to 107,
respectively, for HD~80731, while for HD~17310, which has particularly low
photometric amplitudes and suspiciously high radial-velocity ones, the
ratios are all very large, between 170 and 430.  We suppose that the
observed ratios will be of significance for modellers of the \gds
phenomenon.

Stars in binary systems with short orbital periods often have synchronized
rotations.  (They often have circular orbits, too, but since the time-scale
for circularization is much longer than that for capture of the
rotation$^{20,21}$, the non-zero orbital eccentricities of the three stars
with which we are concerned here does not necessarily imply that the
rotations are not synchronized.)  We have seen in the introductory section
of this paper that all three stars have colours and luminosities appropriate
to early-F dwarfs, so they must also have normal radii of about 0.9 Gm.  The
pseudo-synchronous$^{22}$ rotation periods appropriate to the periods and
orbital eccentricities of the three stars are about 23, 8.2, and 5.2 days,
corresponding to equatorial rotational speeds of about 3, 8, and 13 \kmss
for HD~17310, HD~70645, and HD~80631, respectively.  The variability of the
observed line-widths of all three stars demonstrates that those widths are
not to be interpreted purely in terms of rotation; we do not know whether
the {\it minimum\/} width observed for each is or is not still largely
increased from the value set mainly by rotation, but it must represent an
upper limit to the rotational velocity.  The minimum observed value is very
likely to have been minimized partly by accidental observational error;
allowing subjectively for such an effect, we might say that the minimal
values for the three stars are about 6, 9, and 10 \kms , respectively.
Comparison with the values calculated on the basis of pseudo-synchronism
leads to the conclusions that HD~17310 and HD~70645 are either rotating
faster than synchronism or else their minimum line-widths still owe
something to pulsation, whereas HD~80731 is either rotating more slowly than
synchronism or, if synchronized, has an orbital inclination no greater than
about 50\dego .

\refs\ref1 A. B. Kaye \setal \pasp{111} 840, 1999.
\ref2  G. Handler, \mn{309} L19, 1999.
\ref3  R. F. Griffin, \obs{126} 119, 2006\hm1(Paper 187).
\ref4  K. Krisciunas \setal \mn{273} 662, 1995.
\ref5  F. M. Zerbi \setal \mn{290} 401, 1997.
\ref6  A. B. Kaye, R. O. Gray \& R. F. Griffin, \pasp{116} 558, 2004.
\ref7  S. Martin, M. Bossi \& F. M. Zerbi, \aa{401} 1077, 2003.
\ref8  P. Mathias \setal \aa{417} 189, 2004.
\ref9  G. W. Henry, F. C. Fekel \& S. M. Henry, \aj{129} 2815, 2005.
\r{10} S. Grenier \setal \aas{137} 451, 1999.
\r{11} [Announced by] C. Koen \& L. Eyer, \mn{331} 45, 2002.
\r{12} B. Nordstr\"om \setal \aa{418} 989, 2004.
\r{13} K. Topka \setal \apj{259} 677, 1982.
\r{14} J. H. Moore \& G. F. Paddock, \apj{112} 48, 1950.
\r{15} D. Gillet \setal \aas{108} 181, 1994.
\r{16} R. F. Griffin, R. P. Church \& C. A Tout, {\sit New Astr.}, {\sbf 11},
       431, 2006.\hm1(See {\sit Appendix}.)
\r{17} {\sit e.g}.~D. V. Lindley \& W. F. Scott, {\sit New Cambridge Elementary
       Statistical Tables\/} (CUP), 1984, pp.~50--54.
\r{18} C. L. Morbey \& R. F. Griffin, \apj{317} 343, 1987.
\r{19} R. F. Griffin, \obs{126} 1, 2006\hm1 (Paper 186).
\r{20} J. P. Zahn, \aa{57} 383, 1977; {\sbf 220}, 112, 1989.
\r{21} J.-L. Tassoul, \apj{324} L71, 1988; {\sbf 358}, 196, 1990.
\r{22} \hut

\figsnp\fig1  Radial-velocity traces of HD~80731, obtained with the
Cambridge \smcors on 2006 May 9 (left) and June 3, illustrating the
variability of the `dip' profile.

\fig2  \f{HD 70645}  Cambridge observations are represented by the filled
squares; those published by Henry \smea\up9 are plotted as open circles.
All were give equal weight in the solution of the orbit.

\fig3  Illustrating the most convincing pulsational period detected in the
orbital radial-velocity residuals of HD~70645.  The data points are the
times of observation and the velocity residuals tabulated for the Cambridge
observations in the fifth column of Table~I.  The Henry \smeas velocities
cannot usefully be plotted, because the pulsational period is not determined
well enough to maintain phases back to previous seasons.

\fig4  As Fig.~2, but for HD~80731.  The errant open-circle point below the
maximum of the velocity curve was omitted from the solution of the orbit.

\fig5  \cl{As Fig.~2, but for HD~17310.}

\fig6  Illustrating one of the apparent periods noticed in the
radial-velocity residuals from the orbit of HD~17310 but (it seems) not in
the photometry.  The period was first noticed in the Cambridge observations
alone (the filled squares); if it does not manifest any underlying reality
it is amazing that it should have been reinforced as cogently as it
obviously {\sit is\/} by the inclusion of the altogether independent Henry
\smeas velocities (the open circles).

\end